\documentclass[11pt,a4paper]{article} 
\usepackage[T1]{fontenc} 
\usepackage[utf8]{inputenc} 
\usepackage[english]{babel} 
\usepackage{textcomp} 
\usepackage[english]{varioref} %
\usepackage{float}
\usepackage{comment}
\usepackage{booktabs} 
\usepackage{caption} 
\usepackage{multirow} 
\usepackage{subfig} 
\usepackage{tabularx} 
\usepackage[centertags]{amsmath} 
\usepackage{amsthm} 
\usepackage{amssymb} 
\usepackage{braket} 
\usepackage{tikz} %
\usepackage[numbers, sort&compress]{natbib} 
\usepackage{jheppub} 
\usepackage{todonotes} 
 \usepackage{pgfplots}
\pgfplotsset{compat=1.17}
\usepgfplotslibrary{external}
\usetikzlibrary{pgfplots.groupplots}

\newcommand{\vev}[1]{\langle\!\langle{#1}\rangle\!\rangle}%
\renewcommand{\hat}{\widehat} 
\renewcommand{\tilde}{\widetilde} 
\renewcommand{\bar}{\overline} 
\newcommand{\be}{\begin{equation}} 
\newcommand{\ee}{\end{equation}} 

\numberwithin{equation}{section} 
\newsavebox{\riddlebox} 
	{\begin{lrbox}{\riddlebox}
		\begin{minipage}{\dimexpr\columnwidth-2\fboxsep\relax} 
			\textbf{\Large Questions:} \\ \vspace{0em} \itshape}
		{\end{minipage} 
	\end{lrbox}
	\begin{center} 
  \colorbox{lightgray!40!white}{\usebox{\riddlebox}} 
	\end{center} 
	} 
	{\begin{lrbox}{\riddlebox}
		\begin{minipage}{\dimexpr\columnwidth-2\fboxsep\relax} 
			\textbf{\Large To Do:} \\ \vspace{0em} }
		{\end{minipage} 
	\end{lrbox}
	\begin{center} 
  \colorbox{lightgray!40!white}{\usebox{\riddlebox}} 
	\end{center} 
	} 
\title{Four-point correlation numbers in super Minimal Liouville Gravity in the Ramond sector}
\author{V. Belavin, J. Ramos Cabezas, B. Runov}
\affiliation{Physics Department, Ariel University, 
Ariel 40700, Israel.}

\emailAdd{vladimirbe@ariel.ac.il, juanjose.ramoscab@msmail.ariel.ac.il, borisru@ariel.ac.il}

\abstract{
In this work, we continue the investigation of correlation numbers in $\mathcal{N}=1$ super Minimal Liouville Gravity (SMLG), with physical fields  in the Ramond sector. Building upon our previous construction of physical operators and the evaluation of three-point correlation functions involving Ramond and Neveu-Schwarz (NS) insertions, we now turn to the analytic computation of four-point correlation numbers. This development is motivated by the framework established for the bosonic Minimal Liouville Gravity and its supersymmetric NS analog, where the integration over moduli space in correlation functions can be performed explicitly using the higher equations of motion (HEM) in Liouville theory. In particular, if one of the insertions corresponds to a degenerate field, the four-point amplitude can be expressed in terms of boundary contributions obtained from the OPE structure of logarithmic counterparts of ground ring elements. We aim to adapt and generalize this approach to the Ramond sector. Our result is a closed-form analytic expression for four-point correlation numbers involving Ramond fields.}
\begin{document}
\maketitle

\section{Introduction}
Minimal Liouville Gravity (MLG) and its $\mathcal{N}=1$ supersymmetric extension, Super Minimal Liouville Gravity (SMLG), are exactly solvable laboratories for two-dimensional quantum gravity coupled to conformal matter. In bosonic MLG and in the NS sector of SMLG, analytic four-point correlation numbers were obtained using the Higher Equations of Motion (HEM) of Liouville theory~\cite{Belavin:2005yj,Belavin:2008vc,Belavin:2009ag}. For degenerate insertions, these relations reduce the moduli integral to boundary terms fixed by OPEs of logarithmic ground-ring operators with the remaining physical fields.

The analogous Ramond-sector analysis is still incomplete. Although it has long been conjectured that the Ramond sector is captured by the same discrete matrix-model framework~\cite{Seiberg:2003nm}, explicit moduli-integrated correlators with Ramond insertions have not been derived in the continuous approach. In our previous work, we solved the three-point problem by constructing Ramond physical operators $\mathbb{R}_a$ and evaluating $\langle \mathbb{R}_{a_1}\mathbb{R}_{a_2}\mathbb{W}_{a_3}\rangle$, where $\mathbb{W}_a$ is an NS physical field.

Here we proceed to the next step and compute four-point correlation numbers with Ramond insertions, extending the method of~\cite{Belavin:2008vc,Belavin:2009ag}\footnote{For a close discussion, see \cite{Muhlmann:2025ngz, Rangamani:2025wfa,Eberhardt:2026hfh}.}. Our primary setup is the case where the integrated insertion belongs to the NS sector:
    \begin{equation} \label{4pt1}
     \int d^2z \, \langle \bar{G}_{-1/2}G_{-1/2}\mathbb{U}_{a_1}(z,\bar{z}) \, \mathbb{R}_{a_2}(z_2, \bar{z}_2)\, \mathbb{R}_{a_3}(z_3,\bar{z}_3) \, \mathbb{W}_{a_4}(z_4,\bar{z}_4) \rangle \; ,
    \end{equation}
where $\mathbb{R}_{a}=\mathbb{U}^{R}_{a}\bar{c}c\bar{\sigma}\sigma$ is the BRST-invariant Ramond physical field and $\mathbb{W}_{a}$ is the NS physical field in picture $-1$. The remaining insertion is the non-local physical field obtained from the NS operator in picture $0$, and $G_{-1/2}$ denotes the matter and Liouville part $G^{\text{M+L}}_{-1/2}$ of the supercurrent mode. The picture numbers are therefore $-1/2$ for each Ramond field, $-1$ for $\mathbb{W}_{a_4}$ and $0$ for the integrated insertion, adding up to $-2$ in each chiral sector, as required on the sphere. We derive an explicit expression for~(\ref{4pt1}) below.  The parameter $a_1$ considered here corresponds to the special values labelled by two integers $(m,n)$, with Liouville momentum parameter $a_1=a_{m,-n}=(1-m)/(2b)+(1+n)b/2$, so that the corresponding matter momentum $a_1-b$ is associated with a degenerate matter field. The OPE analysis required for the derivation of (\ref{4pt1}) is carried out explicitly for the simplest case $(m,n)=(1,3)$, for which $a_1=a_{1,-3}=2b$ and therefore the corresponding matter momentum is $a_1-b=b$.  The extension of the resulting structure to generic labels $(m,n)$ is not derived explicitly here and should therefore be regarded as a conjectural generalization, motivated by the explicit $(1,3)$ computation and by the analogous structure known in the NS sector. 

We also comment on the complementary configuration in which the integrated insertion is Ramond; that case contains an important subtlety that is not yet fully resolved.
In both configurations, once one insertion is degenerate, the integrand can be written as a total derivative modulo BRST-exact terms, so the moduli integral is reduced to boundary data.

The paper is organized as follows. In section~\ref{superMLG} we summarize the SMLG setup and notation. Section~\ref{secPhys} presents the NS and Ramond physical operators, together with ghost-number constraints and ground-ring ingredients. We then  in section \ref{a4ptf} derive the required OPE data and evaluate the corresponding four-point correlation numbers. In the final sections \ref{dothercn}  and \ref{conclusions} we discuss implications, open questions and directions. Appendix \ref{appsa} is added as a complement to section \ref{superMLG}.

\section{Super MLG}
\label{superMLG}
In this section we fix notation and summarize standard ingredients of SMLG; additional known formulas are collected in Appendix \ref{appsa}. The SMLG model~\cite{David:1988hj,Distler:1988jt,Distler:1989nt} is a tensor product of superconformal matter (SM), super
Liouville \cite{Polyakov:1981re, curtright1984weak, arvis1983spectrum, d1983classical, babelon1984monodromy}, and super ghost systems \cite{Friedan:1985ge, verlinde1989lectures, Polchinski:1998rr, friedan2003tentative}, with the action
\begin{equation}
A_{\text{SLG}}=A_{\text{SM}}+A_{\text{SL}}+A_{\text{SG}},
\end{equation}
each of which obeys the symmetry~\eqref{N1algebra} with  the central
charge parameters constrained  by
\begin{equation}
\hat{c}_{\text{SM}}+\hat{c}_{\text{SL}}+\hat{c}_{\text{SG}}=0\;. \label{totc}
\end{equation}
Each component theory realizes the same $\mathcal{N}=1$ superconformal symmetry, with both NS and Ramond sectors.

In the continuous formulation we use an extension of minimal Liouville gravity in which the matter part is a generalized minimal model with non-rational $\hat{c}_{SM}$ and a spectrum containing non-degenerate primaries. To match with matrix-model observables, one eventually imposes the usual constraints on central charges and conformal dimensions.
\subsection{Liouville sector}
The Liouville central charge is parameterized by a coupling $b$:
\be
\hat{c}_L=1+2Q^2\,, \quad  Q=b^{-1}+b.
\ee
Primary fields of super Liouville theory are exponential operators. For NS bosonic fields $V_a$ and Ramond fields $R_a^{\epsilon}$:
\begin{equation} \label{SLprimaries}
    V_a(z)= e^{a \varphi (z)},  \quad R_{a}^{\epsilon}(z)= \Sigma^{\epsilon}  e^{a \varphi(z)},
\end{equation}
where $\Sigma^{\epsilon}$ denotes the twist field. The conformal weights of $V_a$ and $R^{\pm}_a$ are
\be
\Delta^{L,NS}(a)=\frac{a(Q-a)}{2}\,,\quad \Delta^{L,R}(a)=\frac{1}{16}+\frac{a(Q-a)}{2}.
\ee
Except at $\Delta=\hat{c}/16$, the Ramond highest-weight modules are doubly degenerate. The two level-zero states are exchanged by the fermionic zero mode $G_0$:
\be
\label{betaL}
    G^{L}_{0}|R^{\sigma}_a\rangle=i\beta^{L}_{a}e^{-\frac{i\pi\sigma}{4}}|R^{-\sigma}_a\rangle\,,\quad
    \beta^{L}_a=\frac{1}{\sqrt{2}}\left(\frac{Q}{2}-a\right)\,.
\ee

Liouville structure constants used later are summarized in Appendix \ref{appsa}.

\subsection{Matter sector}
In SMLG, conformal matter is described by a supersymmetric generalized minimal model (GMM). Unlike rational minimal models, its central charge is allowed to be non-rational. NS primaries are denoted by $\Phi_{\alpha}$ (with continuous parameter $\alpha$), and Ramond primaries by $\Theta_{\alpha}^{\pm}$.
The GMM can be related to super Liouville theory through the analytic continuation
\be
\label{subsML}
    b\to  ib\,,\quad a \to -i a\,,
\ee
with a different normalization chosen so that the two-point functions are unity:
\be \label{matternormalization}
   \langle \Psi_{\alpha} \Psi_{\alpha}\rangle= G_{NS}^{M}(\alpha)=G_{R}^{M} (\alpha)=1\,.
\ee
Here $\Psi_{\alpha}$ stands for a matter primary. Degenerate primaries are indexed by integer pairs $(m,n)$:

\be
    \Phi_{n,m}=\Phi_{\alpha_{n,m}}\,,\quad
    \Theta_{n,m}^{\pm}=\Theta_{\alpha_{n,m}}^{\pm}\,,
    \quad \alpha_{n,m}=\frac{q}{2}+\lambda_{-m,n},\,
\ee
where $m-n$ is even for NS fields and odd for Ramond fields. Condition~\eqref{totc} fixes the matter central charge in terms of $b$:
\be
\hat{c}_M=1-2q^2\,, \quad  q=b^{-1}-b.
\ee
The corresponding NS and Ramond conformal dimensions are
\be
\Delta^{M,NS}(\alpha)=\frac{\alpha(\alpha-q)}{2}\,,\quad \Delta^{M,R}(\alpha)=\frac{1}{16}+\frac{\alpha(\alpha-q)}{2}.
\ee
The normalization of $\Theta^{-}_{\alpha}$ is chosen so that the matter coefficient $\beta^{M}_\alpha$ entering the action of $G_0^{M}$ is tied to its Liouville analog $\beta^{L}$:
\be
\label{betaM}
    G_0^{M}|\Theta_{\alpha}^{\sigma}\rangle=
    i\beta^{M}_{\alpha}e^{-\frac{i\pi\sigma}{4}}
    |\Theta_{\alpha}^{-\sigma}\rangle\,,\quad
    \beta^{M}_{a-b}=-i\beta^{L}_a\,.
\ee
The special matter structure constants follow from the special Liouville ones by applying (\ref{subsML}) together with the normalization (\ref{matternormalization}); explicit expressions are given below.
\subsection{Ghost sector}
The super-ghost theory $A_{\text{SG}}$ is a free superconformal system with central charge $\hat{c}_{gh}=-10.$
It contains anticommuting fermionic ghosts $(b,c)$ with spins $(2,-1)$ and bosonic ghosts $(\beta,\gamma)$ with spins $(3/2,-1/2)$. The field $\delta(\gamma(0))$ (of dimension $1/2$) enters in NS physical fields and amplitudes~\cite{Belavin:2008vc} discussed below. The fundamental OPEs of basic fields are
\begin{equation} \label{bcgbetaope}
    b(z)c(0) =\frac{1}{z}, \quad \gamma(z) \beta(0) = \frac{1}{z}.
\end{equation}
The modes of the fermionic ghosts $b_n,c_n$ are labeled by integers.
The bosonic modes $\beta_k,\gamma_k$ are half-integer moded in the NS sector and integer moded in the Ramond sector. The corresponding super-Virasoro generators are
\begin{align}
\label{ghost-Vir-gener}
&L_m^{g}=\sum_n(m+n):b_{m-n} c_n: +\sum_k(\frac{m}2+k):\beta_{m-k} \gamma_k:  + a^{gh} \delta_{m,0}  ,\\
&G_k^{g}=-\sum_n\left[(k+\frac{n}2):\beta_{k-n} c_n:+2 b_n\gamma_{k-n}\right],
\end{align}
where $a^{gh}=-1/2$ in the NS sector and $a^{gh}=-5/8$ in the R sector.

In the Ramond sector we also use the vacuum state $|v,-1/2\rangle$ in picture $q=-1/2$, with corresponding field $\sigma(z)$ of conformal dimension $3/8$. In the bosonized representation $\sigma \sim e^{-\phi/2}$. We will also use the companion field $\sigma_2 \sim e^{\phi/2}$, with conformal dimension $-5/8$.

\hfill\break
\textbf{Ghost counting rules}
\hfill\break
References~\cite{Belavin:2008vc, Belavin:2025rtg} show that nontrivial correlators must satisfy the following field-counting rules
\begin{equation} \label{gbalance1}
    N_c-N_b=3,
\end{equation}
\begin{equation} \label{gbalance2}
        N_{\delta(\gamma)} - N_{\delta (\beta)}+ N_{\beta}- N_{\gamma}=2.
\end{equation}
Including $\sigma$ and $\sigma_2$ insertions in the Ramond sector modifies the second condition to
\begin{equation}
\label{Nbg_gen}
     N_{\delta(\gamma)} - N_{\delta (\beta)}+ N_{\beta}- N_{\gamma} + N_{\sigma}/2- N_{\sigma_2}/2 =2.
\end{equation}
If only $\delta(\gamma)$, $\sigma$, and $\sigma_2$ appear in the $\beta\gamma$ sector, this reduces to
\begin{equation} \label{sigmarule}
         N_{\delta(\gamma)} + N_{\sigma}/2- N_{\sigma_2}/2 =2.
\end{equation}

\section{Physical fields}\label{secPhys}
Physical fields in SMLG are selected by BRST closure with respct to the charge $\mathcal{Q}$\footnote{The BRST charge $\mathcal{Q}$ in (\ref{PhysCond}) can be written as\begin{align*}
\mathcal{Q}=\sum_m{:}\bigg[L_m^{\text{M+L}}+\frac{1}{2}L^{\text{g}}_m\bigg]c_{-m}{:}+
\sum_r{:}\bigg[G_r^{\text{M+L}}+\frac{1}{2}G^{\text{g}}_r\bigg]\gamma_{-r}{:}+\frac{a^{gh}}{2}c_0.
\end{align*}} together with vanishing total conformal dimension. These requirements can be summarized as
\begin{align}
\label{PhysCond}
    &\mathcal{Q}|\Psi\rangle=0,\quad |\Psi\rangle\neq \mathcal{Q} [...],\\
&L_0|\Psi\rangle=0.
\end{align}

\subsection{Physical fields in the NS sector}
In the NS sector there are two basic classes of physical fields \cite{Belavin:2008vc}:
\begin{align} 
\mathbb{W}_{a}(z,\bar z)=\mathbb{U}_{a}(z,\bar z)\cdot c(z)\bar c(\bar
z)\cdot \delta(\gamma(z))\delta(\bar\gamma(\bar z)), \label{W}
\end{align}
and
\begin{align} 
\tilde{\mathbb{W}}_{a}(z,\bar z)=\biggl(\bar G^{\text{M+L}}_{-1/2}+
\frac12\bar G_{-1/2}^{\text{g}}\biggr)\biggl(G^{\text{M+L}}_{-1/2}+
\frac12G_{-1/2}^{\text{g}}\biggr)\mathbb{U}_{a}(z,\bar z)\cdot\bar
c(\bar z)c(z), \label{tildeW}
\end{align}
where
\begin{align}
\mathbb{U}_{a}(z,\bar z)=\Phi_{a-b}(z,\bar z)V_{a}(z,\bar z).
\end{align}
The momentum parameter $a$ is generic.

Besides these continuous operators, the NS spectrum contains ``discrete states'' built from ground-ring fields:
\begin{equation} \label{opensomn}
\mathbb{O}_{m,n}(z,\bar z)=\bar H_{m,n}H_{m,n}\Phi_{m,n}(z,\bar
z)V_{m,n}(z,\bar z).
\end{equation}
The operators $H_{m,n}$ are polynomials in super-Virasoro generators, defined uniquely up to $\mathcal{Q}$-exact terms.

For analytic computations, two key identities are
\begin{equation}
  \bar{\mathcal{Q}} \mathcal{Q} \mathbb{O}'_{m,n}=B_{m,n}\tilde{\mathbb{W}}_{m,-n} \label{basic},
\end{equation}
and
\begin{align}
\bar G_{-1/2}G_{-1/2}\mathbb{U}_{m,-n}=
B_{m,n}^{-1}\bar\partial\partial \mathbb{O}'_{m,n}\mod \mathcal{Q},
\label{GGU1}
\end{align}
where the logarithmic counterparts of the discrete states
$\mathbb O_{m,n}$,
\begin{equation} \label{opelog1}
\mathbb{O}'_{m,n}=\bar H_{m,n}H_{m,n}\Phi_{m,n}V_{m,n}',
\end{equation}
and $B_{m,n}$ are the coefficients arising        from the higher equations of motion (\ref{HEM}) of SLFT~\cite{Belavin:2006pv}. In what follows we focus on the illustrative case $\mathbb{O}_{13}$:
\begin{equation}  \label{deri3pt3}
\begin{aligned}
\mathbb{O}_{13}(z) =\, & \Phi_{13}^{\prime}(z)\, V_{13}(z)
- \Phi_{13}(z)\, V_{13}^{\prime}(z) -
\Psi_{13}(z)\, \Lambda_{13}(z) +\\
& + \left[b^2 : \beta(z)\gamma(z): + 2 b^2 : b(z) c(z): \right] \Phi_{13}(z)\, V_{13}(z)
 \\
& - b^2 \beta(z)\, c(z)\, \Psi_{13}(z)\, V_{13}(z)
- b^2 \beta(z)\, c(z)\, \Phi_{13}(z)\, \Lambda_{13}(z),
\end{aligned}
\end{equation}
where $\Lambda_{13}=G^{L}_{-1/2}V_{13}$ and $\Psi_{13}=G^{M}_{-1/2}\Phi_{13}$.

{\bf Non-local fields}. Besides the local physical fields, we will also need non-local BRST-invariant
fields of the form
\begin{equation} \label{nonlocalF}
   \int d^2z\,\bar b_{-1}b_{-1}\Psi(z),
\end{equation}
where $\mathcal{Q}\Psi=0$. Indeed, the BRST invariance follows from the fact that 
$ \mathcal{Q} b_{-1}\Psi=L_{-1}\Psi \mod \mathcal{Q}$. By applying this idea to the field (\ref{tildeW}), one obtains the integrated field \( \int d^2z \, \bar{G}_{-1/2}G_{-1/2}\mathbb{U}_{a_1}(z,\bar{z})\) in (\ref{4pt1}), where only $G^{\text{M+L}}_{-1/2}$ survives, because on the picture-zero ground state the ghost part of $G_{-1/2}$ is proportional to $b_{-1}$ and is annihilated by the operators in (\ref{nonlocalF}).

\subsection{Physical fields in the R sector}

Background on the Ramond sector of SMLG can be found in~\cite{Belavin:2025rtg, Distler:1989nt}. Physical states $|\Psi\rangle$ satisfy the cohomological constraints above, and in the Ramond sector one also imposes $\beta_0|\Psi\rangle=G_0|\Psi\rangle=0.$ The physical field in the Ramond sector is therefore given by
\be
\label{RamondPhys}
    \mathbb{R}_a=\mathbb{U}_a^{R}\bar{c}c\bar{\sigma}\sigma,
\ee
where \be \label{URoption2}
\mathbb{U}^{R}_a=\Theta_{a-b}^{-}R_{a}^{+}+i\Theta_{a-b}^{+}R_a^{-}.
\ee
Ramond ``discrete states'' are constructed as
\be \label{groundringop1}
    \mathbb{O}_{m,n}=\bar{H}_{m,n}H_{m,n} \mathbb{Y}^{R}_{m,n}\bar{\sigma}\sigma\,,\quad \mathbb{Y}^{R}_{m,n}=\left(\Theta_{m,n}^{-}R_{m,n}^{+}+i\Theta_{m,n}^{+}R_{m,n}^{-}\right)\,,
\ee
where $H_{mn}$ has dimension $\frac{mn}{2}-1$ and ghost numbers $N_c=0$, $N_{\sigma}=0$, and is built from matter/Liouville super-Virasoro generators and ghosts. For the simplest Ramond ground-ring element one finds \cite{Belavin:2025rtg} 
\be \label{groundringop3}
    H_{1,2}=\frac{1}{2}+\frac{b^2}{1-2b^2}G_0^{M}\beta_{-1}c_1-\frac{b^2}{1+2b^2}G_0^{L}\beta_{-1}c_1+\frac{4b^2}{(1-2b^2)(1+2b^2)}G_0^{L}G_0^{M}.
\ee
One can further show that physical fields satisfy
\be
\label{RQQrep}
\bar{\mathcal{Q}}\mathcal{Q}\mathbb{O}_{m,n}^{\prime}=B_{m,n}\mathbb{R}_{m,-n}.
\ee

\section{Analytical four-point function} \label{a4ptf}
The goal of this section is to derive analytically the four-point correlation number introduced in (\ref{4pt1}). Following the same logic as in the NS case, the moduli integral can be reduced to boundary terms, and these boundary terms are fixed by OPEs of the logarithmic ground-ring operator with the physical fields insertions.

\subsection{Strategy and required OPE data}
Expression (\ref{4pt1}) requires the OPEs of $\mathbb{O}_{1,3}$ with all physical fields entering the correlator. We organize the required OPE data as follows:
\begin{equation}
\begin{aligned}
\mathbb{O}_{1,3}\,\mathbb{W}_{a} &= \sum_{\eta=-1,0,1} A^{NS}_{a+\eta b}\,\mathbb{W}_{a+\eta b},\\
\mathbb{O}_{1,3}\,\widetilde{\mathbb{W}}_{a} &= \sum_{\eta=-1,0,1} \widetilde{A}^{NS}_{a+\eta b}\,\widetilde{\mathbb{W}}_{a+\eta b},\\
\mathbb{O}_{1,3}\,\mathbb{R}_{a} &= \sum_{\eta=-1,0,1} A^{R}_{a+ \eta b}\,\mathbb{R}_{a+\eta b}.
\end{aligned}
\label{genope13}
\end{equation}
The NS OPE data $A^{NS}_{a+\eta b}, \widetilde{A}^{NS}_{a+\eta b}$ have been computed before \cite{Belavin:2008vc}. Here we only need their counterparts for the logarithmic operator $\mathbb{O}'_{1,3}$. The genuinely new ingredient is the OPE $\mathbb{O}_{1,3}\mathbb{R}_a$, which controls the Ramond contribution to the boundary terms. The expected physical structure is diagonal in the BRST cohomology:
\begin{equation} 
    \mathbb{O}_{1,3} \mathbb{R}_a= \sum_{\eta =1,-1,0} A^R_{a+ \eta b}   \mathbb{R}_{a+\eta b}.
\label{targetOPE13R}
\end{equation}
We now present the explicit computation.

\subsection{Special structure constants and auxiliary OPEs}
Here we compute the special structure constants with Ramond fields in the intermediate channel. We extract them from the general expression (\ref{Ramondstructconst}) by taking the appropriate degenerate limits, and we use the resulting constants in the next subsection to determine the OPE coefficients in (\ref{targetOPE13R}).

\subsection{$\mathbb{O}_{1,3}\mathbb{R}_a$}
Here we compute the OPE $\mathbb{O}_{1,3} \mathbb{R}_a$. We take $\mathbb{O}_{1,3}$ from (\ref{deri3pt3}) and aim to reproduce (\ref{targetOPE13R}). To compute the OPE coefficients $A_{a + \eta b}$, we need the special structure constants (\ref{Ramondstructconst}) at the degenerate value $a_3=a_{1,3}=-b$. This can be done by substituting in (\ref{Ramondstructconst})
\begin{equation} \label{specialvalues1}    a_1=a, \quad  a_2=Q- (a+\eta b), \quad a_3=-b+x,
\end{equation}
and computing the limiting expression $x \to 0$ (see appendix B of \cite{Belavin:2007gz} for a more detailed explanation). The result we obtain is as follows:
\begin{equation} \label{clpb}
       \mathbf{C}_{(-b) [a]}^{   [a +  b], L, \epsilon     }= \frac{\epsilon  \left(2 \pi ^2 \mu^2 \gamma \left(\frac{1}{2} \left(b^2+1\right)\right)^2 \gamma \left(b \left(a-\frac{b}{2}\right)\right) \gamma \left(\frac{2 a b-1}{2 b^2}\right)\right)}{\gamma \left(b \left(a+\frac{b}{2}\right)\right) \gamma \left(\frac{a+b-\frac{1}{2 b}}{b}\right)},
\end{equation}
\begin{equation} \label{clmb}
       \frac{\mathbf{C}_{(-b) [a]}^{   [a -  b], L, \epsilon     }}{2}=  \epsilon,
\end{equation}

\begin{equation} \label{cl0b}
     \mathbf{C}_{(-b) [a]}^{   [a ], L, \epsilon     } =    \frac{2 \pi  b^2 \mu \gamma \left(b (a-b)-\frac{1}{2}\right)}{\gamma \left(-b^2\right) \gamma \left(a b+\frac{1}{2}\right)}.
\end{equation}
We will also need the corresponding matter special structure constants. For this we perform the transformations $a \to -i a , b \to i b$, divide the Liouville structure constants by $\sqrt{G_R}$ and $\sqrt{G_{NS}}$, and introduce the proper normalization. As an illustration, from (\ref{clpb})--(\ref{cl0b}) we obtain the relation
\begin{equation}
    \left.    \mathbf{C}_{ (b)[a]}^{[a + \eta b], M, \epsilon}=  \frac{\mathbf{ C}_{ (-b)[a]   }^{[a - \eta b], L, \epsilon}   }{ f_0(b) (G_{NS}(-b) G_{R}(a) G_{R} (Q- (a -\eta b)))^{1/2}} \right|_{\substack{a \to -i a \\ b \to i b}}
,
\end{equation}
where $f_0(b)$ is chosen according to the normalization (\ref{matternormalization}), that is $\mathbf{C}_{ [b/2][b/2]}^{(0), M, \epsilon}=1$. Notice also that in the denominator the argument of $G_{R}$ is shifted by $Q- (a - \eta b)$, since it corresponds to the upper index. We obtain the results:
\begin{equation} \label{cmpb}
     \mathbf{C}_{(b) [a]}^{   [a +  b], M, \epsilon     }= -\frac{\epsilon  \left( \gamma \left(-\frac{3}{2} \left(b^2-1\right)\right) \gamma \left(\frac{1}{2} \left(b^2+1\right)\right) \gamma \left(\frac{a}{b}-\frac{1}{2 b^2}+1\right) \gamma \left(\frac{1-2 b (a+b)}{2 b^2}\right)\right)^{1/2}}{b^2 \left(\gamma \left(\frac{1}{2} b (2 a+b)\right) \gamma \left(-a b-\frac{3 b^2}{2}+1\right)\right)^{1/2}},
\end{equation}

\begin{equation} \label{cmmb}
\begin{split}
      \mathbf{C}_{(b) [a]}^{   [a -  b], M, \epsilon     }= & -\frac{\epsilon \left(  \gamma \left(-\frac{3}{2}  \left(b^2-1\right)\right) \gamma \left(\frac{1}{2} \left(b^2+1\right)\right) \gamma \left(\frac{a}{b}-\frac{1}{2 b^2}+1\right) \right)^{1/2}}{b^2 \gamma \left(a b-\frac{b^2}{2}\right) \left( \gamma \left(-a b+\frac{b^2}{2}+1\right) \gamma \left(\frac{1-2 a b}{2 b^2}+1\right)\right)^{1/2} } \times \\&  \times
      \sqrt{\gamma \left(\frac{1}{2} b (2 a+b)\right)} \gamma \left(\frac{1-2 a b}{2 b^2}\right),
\end{split}      
\end{equation}
\begin{equation} \label{cm0b}
     \mathbf{C}_{(b) [a]}^{   [a ], M, \epsilon     }= \frac{\gamma \left(\frac{1}{2} \left(b^2+1\right)\right) \gamma \left(\frac{1}{2}-a b\right) \gamma \left(b (a+b)-\frac{1}{2}\right)}{ \gamma \left(b^2\right) \left( \gamma \left(\frac{3}{2}-\frac{1}{2 b^2}\right) \gamma \left(\frac{1}{2} \left(\frac{1}{b^2}-1\right)\right)  \gamma \left(\frac{1}{2}-\frac{b^2}{2}\right) \gamma \left(\frac{3 b^2}{2}-\frac{1}{2}\right) \right)^{1/2}}.
\end{equation}
We will also require the following special structure constants:
\begin{equation}
\begin{split}
    & \tilde{\mathbf{C}}_{(-b) [a]}^{   [a + \eta b], L, \epsilon    }   ,  \tilde{\mathbf{C}}_{(b) [a]}^{   [a+ \eta b ], M,  \epsilon   } , \\
    &\mathbf{ d}_{(-b) [a]}^{[a+ \eta b],L, \epsilon }  , \mathbf{d}_{(b) [a]}^{[a+ \eta b], M, \epsilon } .
\end{split}
\end{equation}
These are computed by substituting the special values (\ref{specialvalues1}) in the definitions (\ref{tildeandd}) and using the expressions (\ref{clpb}-\ref{cl0b}) and (\ref{cmpb}-\ref{cm0b}). 
With these data at hand, one can compute the OPE coefficients $A_{a+\eta b}^R$. The terms that provide nontrivial contributions are:
\begin{equation}
\mathbb{O}_{13}(z) = (\hat{O}_1+ \hat{O}_2)  (\hat{\bar{O}}_1+ \hat{ \bar{O}}_2) \Phi_{13} (z)V_{1,3}(z)+... \quad,
\end{equation}
\begin{equation}
\begin{split}
  &  \hat{O}_1= L_{-1}^M-L^L_{-1}+ 2 b^2 b_{-2}c_1 + b^2 \beta_{-1}\gamma_1, \\
  & \hat{O}_2= -G^M_{-1/2} G^L_{-1/2}.
  \end{split}
\end{equation}
Computing the OPE $\mathbb{O}_{1,3} \mathbb{R}_a$, we obtain the coefficients $A^R_{a+ \eta b}$ in the form:

\begin{equation} \label{expforAR1}
\begin{split}
   & A^{R}_{a+ \eta b}=   (\textbf{c}_1-\textbf{c}_2+ \frac{3}{2} b^2)^2 \mathbf{C}_{(b) [a-b]}^{   [a-b+ \eta b ], M, -  } \mathbf{C}_{(b) [a]}^{   [a + \eta b], L, +     }  - \tilde{\mathbf{C}}_{(b) [a-b]}^{   [a-b+ \eta b ], M, -  } \tilde{\mathbf{C}}_{(b) [a]}^{   [a + \eta b], L, +     } \\
   &\quad -i ( \textbf{c}_1-\textbf{c}_2+ \frac{3}{2} b^2 )( \mathbf{d}_{(b) [a-b]}^{[a-b+ \eta b],M,+}  \mathbf{d}_{(-b) [a]}^{[a+ \eta b],L,-}  +  \bar{\mathbf{d}}_{(b) [a-b]}^{[a-b+ \eta b],M,+}  \bar{\mathbf{d}}_{(-b) [a]}^{[a+ \eta b],L,-}),
 \end{split}
\end{equation}
where the coefficients $\textbf{c}_1, \textbf{c}_2$ come from derivatives with respect to the coordinates:
\begin{equation}
\begin{split}
  &  \textbf{c}_1= \Delta^{M,R} (a-b+ \eta b) -   \Delta^{M,R} (a-b) -\Delta^{M,NS} ( b) , \\
  &    \textbf{c}_2= \Delta^{L,R} (a+ \eta b) -   \Delta^{L,R} (a) -\Delta^{L,NS} ( -b) .
\end{split}
\end{equation}
Computing explicitly (\ref{expforAR1}), we arrive at the result:
\begin{equation} \label{finalresultAR1}
    A_{a+ \eta b}^R= K(b) B_{1,3} N_{NS}(2b) \frac{N_R(a)}{N_R(a+ \eta b)},
\end{equation}
where the function $K(b)$\footnote{Our $K(b)$ differs from that in \cite{Belavin:2008vc} by a factor of $1/2$, due to a different normalization of the structure constants.} and the leg factors $N_{NS}$ and $N_R$ are determined as follows:
\begin{equation}
     K(b)=  \frac{1}{b} \left( \gamma \left(\frac{1}{2} \left(\frac{1}{b^2}-1\right)\right) \gamma \left(\frac{1}{2} \left(b^2+1\right)\right) \right)^{1/2},
\end{equation}
\begin{equation} \label{NRlegfactor}
    N_R(a)=\sqrt{\gamma \left(\frac{a}{b}-\frac{1}{2 b^2}\right) \gamma \left(a b-\frac{b^2}{2}\right)} \left(\pi  \mu \gamma \left(\frac{1}{2} \left(b^2+1\right)\right)\right)^{-\frac{a}{b}}\,,
\end{equation}
\begin{equation}
N_{NS}(a)=\left(\pi\mu\gamma\biggl(\frac12+\frac{b^2}2\biggr)\right)^{-a/b}
\left(\gamma\biggl(ab-\frac{b^2}2+\frac12\biggr) \gamma\biggl(\frac
ab-\frac{b^{-2}}2+\frac12\biggr)\right)^{1/2}. \label{N}
\end{equation}
It is worth noting that the form of the result (\ref{finalresultAR1}) is particularly convenient in the computation of the four-point correlation numbers. One can also verify that the remaining terms in the OPE (\ref{targetOPE13R}) which do not contribute to the Ramond physical fields vanish.

We now assemble these data to compute the four-point correlation number.

\subsection{Contour-integral reduction and assembled four-point formula}

Below we proceed with the computation of (\ref{4pt1}). We follow the procedure explained in \cite{Belavin:2005yj,Belavin:2008vc}. For $a_1=a_{1,-3}$, by using relation \eqref{GGU1} and Stokes' theorem, the moduli integral of (\ref{4pt1}) converts into a contour integral:
\begin{equation}
\vev{a_{1,-3}a_2a_3a_4}_{\mathrm{SLG}}
=\frac{1}{B_{1,3}}
\int_{\partial\Gamma}
\partial_z
\left\langle
\mathbb{O}_{1,3}'(z)\,
\mathbb{R}_{a_2}(z_2)\,
\mathbb{R}_{a_3}(z_3)\,
\mathbb{W}_{a_4}(z_4)
\right\rangle
\frac{dz}{2i}.
\label{boundary-master}
\end{equation}
Here
$\partial\Gamma=\partial\Gamma_2+\partial\Gamma_3+\partial\Gamma_4+\partial\Gamma_\infty$, with $\partial\Gamma_i$ around $z_i$ ($i=2,3,4$) clockwise and $\partial\Gamma_\infty$ counterclockwise. As usual, BRST-exact terms do not contribute.

The finite-circle terms are controlled by the logarithmic OPE contributions:
\begin{equation}
\mathbb{O}_{1,3}'(z)\,\frac{\mathbb{W}_{a}(z_4)}{N_{NS}(a)}
=
\log|z-z_4|^2\,
K(b)B_{1,3}N_{NS}(a_{1,-3})
\sum_{r,s\in(1,3)}
q_{r,s}^{(1,3)}(a)\,
\frac{\mathbb{W}_{a+\lambda_{r,s}}(z_4)}{N_{NS}(a+\lambda_{r,s})}
.
\label{logope-q13-NS}
\end{equation}
Analogously, by using the result (\ref{targetOPE13R}, \ref{finalresultAR1}), one can derive a similar expression for the Ramond sector, namely:
\begin{equation}
\mathbb{O}_{1,3}'(z)\,\frac{\mathbb{R}_{a}(z_i)}{N_R(a)}
=
\log|z-z_i|^2\,
K(b)B_{1,3}N_{NS}(a_{1,-3})
\sum_{r,s\in(1,3)}
q_{r,s}^{(1,3)}(a)\,
\frac{\mathbb{R}_{a+\lambda_{r,s}}(z_i)}{N_R(a+\lambda_{r,s})}
,
\quad i=2,3,
\label{logope-q13-R}
\end{equation}
where the coefficients $q_{r,s}^{(1,3)}(a)$ are given by
\begin{equation}
q_{r,s}^{(1,3)}(a)=
\left|a-\lambda_{r,s}-\frac{Q}{2}\right|_{\mathrm{Re}}-\lambda_{1,3},
\qquad
(1,3)=\{(0,-2),(0,0),(0,2)\},
\label{q13def-local}
\end{equation}
with
\begin{equation}
|x|_{\mathrm{Re}}=
\begin{cases}
x, & \mathrm{Re}\,x>0,\\
-x, & \mathrm{Re}\,x<0.
\end{cases}
\label{xredef-local}
\end{equation}
Applying the residue identity (\ref{residue-local}), one computes the finite contour integrals in (\ref{boundary-master}).
\begin{equation}
\frac{1}{2i}\oint_{\partial\Gamma_i}\partial_z\log|z-z_i|^2\,dz=-\pi,
\qquad i=2,3,4.
\label{residue-local}
\end{equation}

The contribution from the infinity contour (curvature contribution) appears because $\mathbb{O}'_{1,3}$ is not a scalar under transformation $z \to y$ :
\begin{equation}
\mathbb{O}_{1,3}'(y)=
\mathbb{O}_{1,3}'(z)-2\Delta'_{1,3}\,\mathbb{O}_{1,3}(z)\log\left|\frac{dy}{dz}\right|,
\label{transform-local}
\end{equation}
\begin{equation}
\Delta'_{1,3}=
\left.
\frac{d}{da}\Delta^{(L)}_{a}
\right|_{a=a_{1,3}}
=\lambda_{1,3}.
\label{delta-prime-local}
\end{equation}
Therefore, as $z\to\infty$,
\begin{equation}
\left\langle
\mathbb{O}_{1,3}'(z)\,
\mathbb{R}_{a_2}(z_2)\,
\mathbb{R}_{a_3}(z_3)\,
\mathbb{W}_{a_4}(z_4)
\right\rangle
\sim
-2\Delta'_{1,3}\log(z\bar{z})
\left\langle
\mathbb{O}_{1,3}\,
\mathbb{R}_{a_2}\,
\mathbb{R}_{a_3}\,
\mathbb{W}_{a_4}
\right\rangle,
\label{asympt-local}
\end{equation}
and
\begin{equation}
\frac{1}{2i}\oint_{\partial\Gamma_\infty}
\partial_z
\left\langle
\mathbb{O}_{1,3}'(z)\cdots
\right\rangle
dz
=
-2\pi\lambda_{1,3}
\left\langle
\mathbb{O}_{1,3}\cdots
\right\rangle.
\label{infinityContribution13}
\end{equation}
Summing the finite boundary terms and the curvature contribution, we obtain
\begin{equation}
\begin{split}
\vev{a_{1,-3}\,a_2\,a_3\,a_4}
=&
\pi K(b) N_{NS}(a_{1,-3})
\left\{
\sum_{i=2}^{4}\sum_{r,s\in(1,3)}q_{r,s}^{(1,3)}(a_i)+6\lambda_{1,3}
\right\}
\\
&\times
\vev{a_2\,a_3\,a_4}_{\mathrm{SLG}}^{(R)},
\end{split}
\label{assembled4pt}
\end{equation}
where $\vev{a_2\,a_3\,a_4}_{\mathrm{SLG}}^{(R)}=\langle \mathbb{R}_{a_2}  \mathbb{R}_{a_3}  \mathbb{W}_{a_4}\rangle$ is the three-point correlation number given by \cite{Belavin:2025rtg}
\begin{equation}
    \vev{a_2a_3a_4}_{\mathrm{SLG}}^{(R)}
=\Omega_R(b)N_R(a_2)N_R(a_3)N_{NS}(a_4),
\end{equation}
where
\begin{equation} \label{ROmegafactor}
  \Omega_R(b)=  \frac{ b^{-3} \left( \pi \mu \right) ^{\frac{1}{b^2}+1} \sqrt{\gamma \left(\frac{1}{2}-\frac{1}{2 b^2}\right)} \gamma \left(\frac{1}{2} \left(b^2+1\right)\right)^{\frac{1}{b^2}+2}  }{\gamma \left(\frac{3}{2}-\frac{1}{2 b^2}\right) \sqrt{\gamma \left(\frac{1}{2} \left(b^2-1\right)\right)}}.
\end{equation}
Using the result (\ref{ROmegafactor}),  one can rewrite (\ref{assembled4pt}) equivalently as
\begin{equation}
\begin{split}
\vev{a_{1,-3}a_2a_3a_4}
=&
\pi K(b)\Omega_R(b)\,
N_{NS}(a_{1,-3})N_R(a_2)N_R(a_3)N_{NS}(a_4)
\\
&\times
\left\{
\sum_{i=2}^{4}\sum_{r,s\in(1,3)}q_{r,s}^{(1,3)}(a_i)+6\lambda_{1,3}
\right\}.
\end{split}
\label{assembled4pt-legfactors}
\end{equation}
Formula (\ref{assembled4pt-legfactors}) represents our main result and corresponds to the direct analogue of the pure NS four-point correlation number found in \cite{Belavin:2008vc}. Two comments are in order. First, one expects that for general ($m,n$) and the parameter $a_{m,-n}$, the formula (\ref{assembled4pt-legfactors}) will hold, so that we have
\begin{equation}
\begin{split}
\vev{a_{m,-n}a_2a_3a_4}
=&
\pi K(b)\Omega_R(b)\,
N_{NS}(a_{m,-n})N_R(a_2)N_R(a_3)N_{NS}(a_4)
\\
&\times
\left\{
\sum_{i=2}^{4}\sum_{r,s\in(m,n)}q_{r,s}^{(m,n)}(a_i)+2 m n\lambda_{m,n}
\right\}.
\end{split}
\label{assembled4pt-legfactors2}
\end{equation}
where in the sum the fusion set is $(m,n)=\{ 1-m:2:m-1, 1-n:2:n-1  \}$. Second, for the purpose of comparing (\ref{assembled4pt-legfactors}, \ref{assembled4pt-legfactors2}) with results from other approaches (e.g. direct computations and matrix model approaches), it is convenient to rewrite the above expressions in a standard normalized form, in which one introduces renormalized fields:
\begin{equation}
\widehat{\mathbb{R}}_a=\frac{\mathbb{R}_a}{N_R(a)},
\qquad
\widehat{\mathbb{W}}_a=\frac{\mathbb{W}_a}{N_{NS}(a)},
\qquad
\widehat{\mathbb{U}}_a=\frac{\mathbb{U}_a}{N_{NS}(a)}.
\label{renorm-fields-R}
\end{equation}
Then the normalized four-point correlation number (\ref{assembled4pt-legfactors}) takes the compact form:
\begin{equation}
\begin{split}
& \frac{1}{\Omega_R(b)}
\int d^2z\,
\left\langle
\bar{G}_{-1/2}G_{-1/2}\widehat{\mathbb{U}}_{1,-3}(z)\,
\widehat{\mathbb{R}}_{a_2}(z_2)\,
\widehat{\mathbb{R}}_{a_3}(z_3)\,
\widehat{\mathbb{W}}_{a_4}(z_4)
\right\rangle
= \\&
\pi K(b)
\left\{
\sum_{i=2}^{4}\sum_{r,s\in(1,3)}q_{r,s}^{(1,3)}(a_i)+6\lambda_{1,3}
\right\}.
\label{renorm-four-point-R}
\end{split}
\end{equation}

\section{Discussion of other type of correlation number} \label{dothercn}
In this section we discuss the complementary four-point correlator of the form\footnote{Notice that the integrated field \( \int d^2 z \,
\mathbb{U}_{1,2}^{R}(z,\bar z)\,\sigma(z)\bar{\sigma}(\bar z) \) in (\ref{alt4pt12}) arises by applying the construction (\ref{nonlocalF}) in the case when the field \( \Psi \) is defined by (\ref{RamondPhys}).}
\begin{equation}
\int d^2 z \,
\left\langle
\mathbb{U}_{1,2}^{R}(z,\bar z)\,\sigma(z)\bar{\sigma}(\bar z)\,
\mathbb{R}_{a_2}(z_2,\bar z_2)\,
\mathbb{W}_{a_3}(z_3,\bar z_3)\,
\widetilde{\mathbb{W}}_{a_4}(z_4,\bar z_4)
\right\rangle .
\label{alt4pt12}
\end{equation}
To apply the contour-reduction logic used in previous section, all the required OPEs in the channels of (\ref{alt4pt12}) must preserve the physical short-distance pattern of Eq.~(\ref{genope13}).

\paragraph{Channel $\mathbb{O}_{1,2}\mathbb{R}_a$.}
This OPE was discussed previously in \cite{Belavin:2025rtg} and was shown to satisfy the required behavior (\ref{genope13}).

\paragraph{Special structure constants.}
In addition, the evaluation of (\ref{alt4pt12}) requires the special structure constants entering the $\mathbb{O}_{1,2}$ analysis. We collect here the relevant formulas:
\begin{equation}
\mathbf{C}_{[-b/2] (a)}^{[a +b /2],L,\epsilon}
=\frac{2 \pi  b^2 \mu \gamma \left(\frac{1}{2} \left(b^2+1\right)\right) \gamma \left(a b-\frac{b^2}{2}-\frac{1}{2}\right)}{\gamma (a b)},
\label{B2}
\end{equation}
\begin{equation}
\frac{\mathbf{C}_{[-b/2] (a)}^{[a -b /2],L,\epsilon}}{2}= \epsilon ,
\label{B3}
\end{equation}

\begin{equation}
\mathbf{C}_{[b/2] (a)}^{[a +b /2],M,\epsilon}
= \epsilon \,
\frac{
\left(
\gamma \left(1-b^2\right)
\gamma \left(\frac{1}{2} \left(b^2-1\right)\right)
\gamma \left(-\frac{2 a b+b^2-1}{2 b^2}\right)
\gamma \left(\frac{2 a b+b^2-1}{2 b^2}\right)
\right)^{1/2}
}{
\left(
\gamma \left(\frac{1}{2} \left(1-\frac{1}{b^2}\right)\right)
\gamma \left(\frac{1}{2} \left(\frac{1}{b^2}-1\right)\right)
\gamma \left(\frac{1}{2} \left(2 a b+b^2-1\right)\right)
\gamma \left(1-b (a+b)\right)
\right)^{1/2}
},
\label{B4}
\end{equation}

\begin{equation}
\begin{split}
\mathbf{C}_{[b/2] (a)}^{[a -b /2],M,\epsilon}
= & -\frac{
b^2
\left(
\gamma \left(1-b^2\right)
\gamma \left(\frac{1}{2} \left(b^2-1\right)\right)
\gamma \left(\frac{-2 a b+b^2+1}{2 b^2}\right)
\gamma \left(\frac{1}{2} \left(2 a b+b^2-1\right)\right)
\right)^{1/2}
}{
\gamma (a b)
\left(
\gamma \left(\frac{1}{2} \left(1-\frac{1}{b^2}\right)\right)
\gamma \left(\frac{1}{2} \left(\frac{1}{b^2}-1\right)\right)
\gamma (1-a b)
\right)^{1/2}
} \times  \\&  \times
\left(\gamma \left(\frac{2 a b+b^2-1}{2 b^2}\right) \right)^{1/2}
.
\label{B5}
\end{split}
\end{equation}

\paragraph{Channel $\mathbb{O}_{1,2}\widetilde{\mathbb{W}}_a$.}
The corresponding OPE is more cumbersome and depends not only on the structure constants above, but also on the associated $\tilde{\mathbf C}$ and $d,\bar d$ structure constants. Nevertheless, the general expectation for this OPE is according to (\ref{genope13}) with the corresponding coefficients $\tilde {A}^R_{a \pm b/2}$ having the form
\begin{equation}
\tilde{A}^R_{a \pm b/2}
= \tilde{f}_{\pm} \frac{K(b)B_{1,2} N_R(a_{1,-2})N_{NS}(a)}{N_{R}(a\pm b/2)} .
\label{B6}
\end{equation}
Organizing the computation term by term, the contributions naturally split into
\begin{equation}
\begin{split}
(\mathrm{I})&=\mathbb{Y}(a_{1,2})\sigma\bar{\sigma}\,\widetilde{\mathbb{W}}_a,\\
(\mathrm{II})&=\left( \frac{b^2}{1-2b^2} G_0^M-\frac{b^2}{1+2b^2}G_0^L \right)\beta_1c_{-1}\,\mathbb{Y}(a_{1,2})\sigma\bar{\sigma}\,\widetilde{\mathbb{W}}_a,\\
(\mathrm{III})&=\left(\frac{b^2}{1-2b^2}\bar{G}_0^M-\frac{b^2}{1+2b^2} \bar{G}_0^L \right)\bar{\beta}_1\bar{c}_{-1}\,\mathbb{Y}(a_{1,2})\sigma\bar{\sigma}\,\widetilde{\mathbb{W}}_a,\\
(\mathrm{IV})&=\left( \frac{b^2}{1-2b^2} G_0^M- \frac{b^2}{1+2b^2}G_0^L\right)\beta_1c_{-1}\left(\frac{b^2}{1-2b^2}\bar{G}_0^M-\frac{b^2}{1+2b^2}\bar{G}_0^L\right)\bar{\beta}_1\bar{c}_{-1}\,\mathbb{Y}(a_{1,2})\sigma\bar{\sigma}\,\widetilde{\mathbb{W}}_a.
\end{split}
\label{B9}
\end{equation}
The resulting intermediate expressions are lengthy. As a representative example, we only provide here the explicit form of the contribution (I):
\begin{equation}
\mathrm{(I)}=\,\left(
\begin{aligned}
&\tilde{\mathbf{C}}^{[a-b\pm b/2],M,-}_{[b/2](a-b)}
\mathbf{C}^{[a\pm b/2],L,+}_{[-b/2](a)}
+\mathbf{C}^{[a-b\pm b/2],M,-}_{[b/2](a-b)}
\tilde{\mathbf{C}}^{[a\pm b/2],L,+}_{[-b/2](a)}
\\
&\quad +\, i\Big(
\bar{ \mathbf{d}}^{[a-b\pm b/2],M,-}_{[b/2](a-b)}\, \mathbf{d}^{[a\pm b/2],L,+}_{[-b/2](a)}
- \mathbf{d}^{[a-b\pm b/2],M,-}_{[b/2](a-b)}\, \bar{\mathbf{d}}^{[a\pm b/2],L,+}_{[-b/2](a)}
\Big)
\end{aligned}
\right)\mathbb{R}_{a\pm b/2}.
\label{B12}
\end{equation}
A direct computation of \eqref{B12} already yields the structure \eqref{B6}, but with a coefficient $\tilde f^{(I)}_{\pm}(a,b)$ rational in $a$ and $b$. The final coefficient $\tilde f_{\pm}$ arises only after summing (I)--(IV); although we do not determine it explicitly here, we expect, by analogy with the previous case, that the $a$-dependence cancels in the sum. This, together with the cancellation of unphysical terms in this channel, requires further investigation.


\paragraph{Channel $\mathbb{O}_{1,2}\mathbb{W}_a$.}
Finally, this channel contains a subtlety in the $\beta\gamma$ ghost sector. In the bosonized language $\sigma\sim e^{-\phi/2}$ and $\delta(\gamma)\sim e^{-\phi}$, hence $\sigma(z)\delta(\gamma)(0)\sim e^{-3\phi/2}(0)$ and is not proportional to $\sigma$ (it shifts the picture from $q=-1/2$ to $q=-3/2$). On the other hand, the correlator $ \langle \sigma\,\delta(\gamma)\,\sigma \rangle $ is non-vanishing. This  problem requires a careful analysis of the picture changing in product of physical fields. For example, in the superstring vertex-operator formalism of \cite{Friedan:1985ge}, the picture-changing operator provides a systematic way to relate physical vertex operators written in different pictures. Its action shifts the picture number in the physical fields. It would therefore be useful to understand whether an analogous picture-changing procedure can be implemented in the present setting and, in particular, whether a physical-field representative in the $q=-3/2$ picture can be obtained through this procedure. We leave this question for future work.

\section{Conclusion} \label{conclusions}
In this work we computed analytically the four-point correlation number (\ref{4pt1}) in super minimal Liouville gravity with Ramond insertions, in the case when the nonlocal physical field corresponds to a degenerate insertion with parameter $a_1=a_{1,-3}$. The key new ingredient is the explicit OPE data for the logarithmic ground-ring operator $\mathbb{O}'_{1,3}$ with Ramond physical states. Combining these OPEs with the higher equations of motion reduction and the logarithmic transformation property (\ref{GGU1}), we obtained a closed-form expression for the four-point correlation number, summarized in (\ref{assembled4pt-legfactors2}).

An important check of this result would be a direct numerical evaluation of the moduli integral in (\ref{4pt1}) and a comparison with the normalized closed-form expression (see (\ref{renorm-four-point-R})). We leave such a numerical test for future work.

It would be very interesting to reproduce the same four-point number from the dual matrix-model description. In particular, after passing to the renormalized fields (\ref{renorm-fields-R}), the normalized correlator (\ref{renorm-four-point-R}) becomes the natural object to compare with matrix-model predictions in the Ramond sector, extending the pure bosonic analysis \cite{Belavin:2005yj} and recent analysis in the supersymmetric case \cite{Johnson:2025vyz}.

Finally, we also discussed the complementary four-point correlator with an integrated Ramond insertion, (\ref{alt4pt12}). While the contour reduction again requires OPE data compatible with the physical pattern (\ref{genope13}), the necessary OPEs appear to be more subtle. We presented the relevant structure constants and the expected behavior of the corresponding OPE coefficients, and we leave a complete analytic evaluation of (\ref{alt4pt12}) for future work.

\appendix

\section{Conformal data}\label{appsa}

\subsection{Superconformal algebra} 
The symmetry algebra of SMLG is $\mathcal{N}=1$ superconformal algebra,  
\begin{equation}
\begin{aligned}
\lbrack L_n,L_m]&=(n-m)L_{n+m}+\frac{\hat c}8(n^3-n)\delta_{n,-m},
\\
\{G_r,G_s\}&=2L_{r+s}+\frac{\hat
c}2\left(r^2-\frac14\right)\delta_{r,-s},
\\
[L_n,G_r]&=\left(\frac12n-r\right)G_{n+r},
\end{aligned}
\label{N1algebra}
\end{equation}
where
\begin{equation*}
\begin{alignedat}{2}
&r,s\:\in\mathbb{Z}+\frac12&\quad&\text{for the NS sector},
\\
&r,s\:\in\mathbb{Z}&\quad&\text{for the R sector}.
\end{alignedat}
\end{equation*}

\subsection{OPE and structure constants}
The basic structure constants in the pure NS sector $\mathbf{C}^{L}_{(a_1)(a_2)(a_3)}$ and
$\tilde{\mathbf{C}}^{L}_{(a_1)(a_2)(a_3)}$ were computed in ~\cite{Poghossian:1996agj,Rashkov:1996np}.
The 3-point functions involving two fields from the Ramond sector and  the superfields from the NS sector can be written as follows \cite{Poghossian:1996agj}
\be \label{apOPERRV}
\begin{split}
    \Big< R^{\epsilon_1}_{\alpha_1}(z_1)R^{\epsilon_2}_{\alpha_2}(z_2)
    \Big(V_{\alpha_3}(z_3)+&\theta G_{-\frac{1}{2}}V_{\alpha_3}(z_3)+\bar{\theta}\,\bar{G}_{-\frac{1}{2}}V_{\alpha_3}(z_3)+\theta\bar{\theta}G_{-\frac{1}{2}}\bar{G}_{-\frac{1}{2}}V_{\alpha_3}(z_3)\Big)\Big>
    \\
    =
  v_{3pt}(z_1,z_2,z_3)
       &
       \left(
    \delta_{\epsilon_1,\epsilon_2} 
    \left(\mathbf{C}^{L,\epsilon_1}_{[\alpha_1][\alpha_2](\alpha_3)}+\frac{|z_{12}|\theta\bar{\theta}}{|z_{13}z_{23}|}\tilde{\mathbf{C}}^{L,\epsilon_1}_{[\alpha_1][\alpha_2](\alpha_3)}\right)\right.
    \\
    &
    \left.
    +\delta_{\epsilon_1,-\epsilon_2}\left(\frac{z_{12}^{\frac{1}{2}}\theta}{(z_{13}z_{23})^{\frac{1}{2}}} \mathbf{d}^{L,\epsilon_1}_{[\alpha_1][\alpha_2](\alpha_3)}+\frac{\bar{z}_{12}^{\frac{1}{2}}\bar{\theta}}{(\bar{z}_{13}\bar{z}_{23})^{\frac{1}{2}}}\,\bar{\mathbf{d}}^{L,\epsilon_1}_{[\alpha_1][\alpha_2](\alpha_3)}\right)
    \right)\\
\end{split}
\ee
with
\be
  v_{3pt}(z_1,z_2,z_3)=|z_{12}|^{-2\gamma_{123}}
     |z_{13}|^{-2\gamma_{132}}
      |z_{23}|^{-2\gamma_{231}},
\ee
\be
    \gamma_{ijk}=\Delta(\alpha_i)+\Delta(\alpha_j)-\Delta(\alpha_k)\,,\quad z_{ij}=z_i-z_j\,.
\ee
With all the structure constants ($\mathbf{C}, \tilde{\mathbf{C}}, \mathbf{d}, \bar{\mathbf{d}}$) are found to be given by \cite{Poghossian:1996agj}
\begin{equation} \label{Ramondstructconst}
\begin{aligned}
\mathbf{C}_{[a_1][a_2] (a_3)}^{L,\epsilon} & =\frac{1}{2}\left(\pi \mu\gamma\left(\frac{Q b}{2}\right) b^{1- b^2}\right)^{\frac{Q-a}{b}} \\
& \times\left[\frac{\Upsilon_{\mathrm{R}}(b) \Upsilon_{\mathrm{R} }\left(2 a_1\right) \Upsilon_{\mathrm{R}}\left(2 a_2\right) \Upsilon_{\mathrm{NS}}\left(2 a_3\right)}{\Upsilon_{\mathrm{R}}(a-Q) \Upsilon_{\mathrm{R}}\left(a_1+a_2-a_3\right) \Upsilon_{\mathrm{NS}}\left(a_2+a_3-a_1\right) \Upsilon_{\mathrm{NS}}\left(a_3+a_1-a_2\right)}\right. \\
& \left.+\epsilon \frac{\Upsilon_{  \mathrm{R} }(b) \Upsilon_{\mathrm{R}}\left(2 a_1\right) \Upsilon_{\mathrm{R}}\left(2 a_2\right) \Upsilon_{\mathrm{NS}}\left(2 a_3\right)}{\Upsilon_{\mathrm{NS}}(a-Q) \Upsilon_{\mathrm{NS}}\left(a_1+a_2-a_3\right) \Upsilon_{\mathrm{R}}\left(a_2+a_3-a_1\right) \Upsilon_{\mathrm{R}}\left(a_3+a_1-a_2\right)}\right].
\end{aligned}
\end{equation}

\begin{align} \label{tildeandd}
\widetilde{\mathbf{C}}^{L, \epsilon}_{[\alpha_{1}],[\alpha_{2}],(\alpha_{3})}
&= i\,\epsilon \left[\left(\beta_1^{2}+\beta_2^{2}\right)
\mathbf{C}^{L,\epsilon}_{[\alpha_{1}],[\alpha_{2}],(\alpha_{3})}
-2\beta_{1}\beta_{2}\,
\mathbf{C}^{L,-\epsilon}_{[\alpha_{1}],[\alpha_{2}],(\alpha_{3})}\right],\\[4pt]
\mathbf{d}^{L,\epsilon}_{[\alpha_{1}],[\alpha_{2}],(\alpha_{3})}
&= i\,e^{-i\pi\epsilon/4}\left[\beta_{2}\,
\mathbf{C}^{L,\epsilon}_{[\alpha_{1}],[\alpha_{2}],(\alpha_{3})}
-\beta_{1}\,
\mathbf{C}^{L,-\epsilon}_{[\alpha_{1}],[\alpha_{2}],(\alpha_{3})}\right], \\
\bar{\mathbf{d}}^{L,\epsilon}_{[\alpha_{1}],[\alpha_{2}],(\alpha_{3})}
&= -i\,e^{i\pi\epsilon/4}\left[\beta_{2}\,
\mathbf{C}^{L,\epsilon}_{[\alpha_{1}],[\alpha_{2}],(\alpha_{3})}
-\beta_{1}\,
\mathbf{C}^{L,-\epsilon}_{[\alpha_{1}],[\alpha_{2}],(\alpha_{3})}\right].
\end{align}
The expressions for $\Upsilon_{\mathrm{NS}}, \Upsilon_{\mathrm{R}}$ can be read from \cite{Poghossian:1996agj, Belavin:2025rtg}. The normalization used in (\ref{Ramondstructconst}) is such that the reflection coefficients (the Liouville two-point functions) are given by
\begin{equation}
\frac{\mathbf{C}_{[a][a_2] (a_3)}^{L,\epsilon}}{\mathbf{C}_{[Q-a][a_2] (a_3)}^{L,\epsilon}}=G_R(a)=\frac{\gamma \left(a b-\frac{b Q}{2}+\frac{1}{2}\right) \left(\pi  \mu \gamma \left(\frac{1}{2} \left(b^2+1\right)\right)\right)^{\frac{Q-2 a}{b}}}{\gamma \left(-\frac{a}{b}+\frac{Q}{2 b}+\frac{1}{2}\right)},
\end{equation}
\begin{equation} \label{refcoefRN}
\frac{\mathbf{C}_{L,[a_1][a_2] (a)}^\epsilon}{\mathbf{C}_{L,[a_1][a_2] (Q-a)}^\epsilon}=  G_{NS}(a)=\frac{b^2 \gamma \left(a b-\frac{b Q}{2}\right) \left(\pi  \mu \gamma \left(\frac{1}{2} \left(b^2+1\right)\right)\right)^{\frac{Q-2 a}{b}}}{\gamma \left(\frac{Q}{2 b}-\frac{a}{b}\right)}\,,
\end{equation}
here we use the notation
\begin{equation} \label{littlegamma}
    \gamma(x)= \frac{\Gamma(x)}{\Gamma(1-x)},
\end{equation}
where $\Gamma(x)$ is the gamma function.

\hfill\break
\textbf{Degenerate fields}
\hfill\break
In our discussion an important role is played by the degenerate fields $V_{m,n}$ and $R^{\pm}_{m,n}$ which one can define as follows
\be
\label{SLdeg}
    V_{m,n}=V_a\Big|_{a=a_{m,n}}\,,\quad
    R^{\pm}_{m,n}=R^{\pm}_a\Big|_{a=a_{m,n}}\,,
    \quad a_{m,n}=\frac{Q}{2}-\frac{b^{-1}m+bn}{2},
\ee
where $m-n$ is even or odd depending on the sector of the fields, as discussed above. From them one introduces logarithmic degenerate fields as follows
\be
    V_{m,n}^{\prime}(z)=\frac{\partial}{\partial a}V_{a}(z)\Big|_{a=a_{m,n}}\,,\quad
    R_{m,n}^{\pm\prime}(z)=\frac{\partial}{\partial a}R^{\pm}_{a}(z)\Big|_{a=a_{m,n}}\,,
\ee
which are subject to Higher Equations of Motion
\be
\label{HEM}
\bar{D}^{L}_{m,n}D^{L}_{m,n}V^{\prime}_{m,n}=B_{m,n}V_{m,-n}\,,\quad
    \bar{D}^{L}_{m,n}D^{L}_{m,n}R^{\pm\prime}_{m,n}=B_{m,n}R^{\pm}_{m,-n}.
\ee
The explicit expressions for factors  $B_{m,n}$ and the operators $D_{m,n}$ can be found in \cite{Belavin:2006pv}.

\clearpage
\bibliographystyle{JHEP} 
\bibliography{bib} 
\end{document}